\documentclass[letterpaper]{article} 
\usepackage{aaai2026}  
\usepackage{times}  
\usepackage{helvet}  
\usepackage{courier}  
\usepackage[hyphens]{url}  
\usepackage{graphicx} 
\usepackage{natbib}  
\usepackage{caption} 
\usepackage{algorithm}
\usepackage{algorithmic}
\usepackage{enumitem}
\usepackage{placeins}
\usepackage{longtable}
\usepackage{booktabs}
\usepackage{url}
\usepackage{newfloat}
\usepackage{listings}
\DeclareCaptionStyle{ruled}{labelfont=normalfont,labelsep=colon,strut=off} 
\floatstyle{ruled}
\newfloat{listing}{tb}{lst}{}
\floatname{listing}{Listing}
\title{A Critical Pragmatism Approach for AI Ethics: Lessons from Urban Planning Theory}
\author {
    Jennah Gosciak\textsuperscript{\rm 1,2},
    Karen Levy\textsuperscript{\rm 2},
    Allison Koenecke\textsuperscript{\rm 1,2}
}
\affiliations {
    \textsuperscript{\rm 1}Cornell Tech\\
    \textsuperscript{\rm 2}Cornell University\\
    jrg377@cornell.edu, karen.levy@cornell.edu, koenecke@cornell.edu
}

\begin{document}

\maketitle

\begin{abstract}
As data scientists grapple with increasingly complex ethical decisions in machine learning (ML) and algorithm design, scholarship on AI ethics has offered multiple solutions, from formal mathematical definitions of fairness to holistic fairness frameworks drawn from various academic disciplines. Still, developing ethical and fair algorithmic systems in practice remains an ongoing challenge. In this paper, we draw a parallel between problems in algorithmic fairness and urban planning. We frame algorithmic fairness problems as `wicked problems,' a term originating from the planning and policy space to describe the intractable, value-laden, and complex nature of this work. As such, we argue that the field of AI ethics can learn from theoretical work in urban planning in ameliorating its own set of wicked problems. Urban planning is typically concerned with practical issues of governance, resource allocation, stakeholder engagement, and conflicts involving deep-seated differences. These are challenges that purely technical approaches in AI easily overlook. We present a flexible set of recommendations based on the urban planning theory approach of \emph{critical pragmatism} --- a reflective and deliberative approach to addressing wicked problems that considers what practitioners actually \emph{do} in the face of conflict and power. We apply our recommendations to several case studies: automated mortgage lending, school choice, and feminicide counterdata collection. Researchers and practitioners can incorporate these recommendations derived from urban planning into their ongoing work to more holistically address problems related to AI ethics.
\end{abstract}


\section{Introduction}

Understanding the social and ethical impact of algorithms is complex and has been the subject of substantial interdisciplinary research with contributions from computer science, law, philosophy, and the social sciences \citep{chouldechovadisparateimpact, kleinberg2016inherent, jacobs2021measurement, binns2018fairness, heidari2019moral, arif2022towards, selbst2019fairness, davis2021algorithmic}. 
In particular, there is a growing body of theoretical work that takes a critical and holistic view of algorithmic fairness -- from data feminism~\citep{d2023data} and non-ideal algorithmic fairness~\citep{fazelpour_algorithmic_2020} to algorithmic realism~\citep{green2020algorithmic} and pipeline-aware fairness \citep{black_toward_2023}.
Implementing the guidance from these holistic approaches in real-world settings, however, remains a challenge. 
Recent work has highlighted the persistent barriers facing industry practitioners and has called for a more pragmatic and practical lens~\citep{holstein2019improving, lee2021landscape, deng2022exploring, smith2025pragmatic, deng2025supporting}.

\textbf{Our work fills this gap by providing a series of practical recommendations that data scientists and engineers can use.} 
We develop these recommendations by comparing \textit{algorithmic fairness problems} to similar problems in \textit{urban planning}.\footnote{We use the term ``urban planning,'' but this literature is also related to policy analysis and public administration~\citep{forester1981questioning}.} Algorithmic fairness problems relate to the complex ethical trade-offs and governance concerns involved in designing algorithmic systems. Planning, broadly, refers to a collection of activities associated with organizing social and political life. Planning problems can be local or global; for example, economic, environmental, and regional planning are also types of planning fields \citep{friedmann1987planning}.

To concretize our analogy, we use the framing of ``wicked problems'' --- a term proposed by planning and design theorists Horst Rittel and Melvin Webber~(\citeyear{rittel_dilemmas_1973}).\footnote{
Though the term wicked problems has been widely cited across numerous fields, both Rittel and Webber were originally speaking to a planning audience. The two worked on research related to design thinking, architecture, and city and regional planning. Horst Rittel was a professor in the Department of Architecture at U.C. Berkeley~\citep{churchman2007memoriam}; Melvin Webber was faculty in the Department of City and Regional Planning at U.C. Berkeley~\citep{webber_bio}.
}
Wicked problems are complicated, value-laden, and subject to reasonable disagreement. The widely influential term has been cited in fields like operations research, software engineering, and human-computer interaction. However, fundamentally, it is a term from the planning literature.

Framing algorithmic fairness problems as ``wicked problems'' allows us to engage in translational work, borrowing ideas from urban planning theorists who have sought to resolve similarly complex ``wicked'' problems for decades. 
Much of the recent theoretical work addressing wicked problems in urban planning has focused on \emph{practical experience} and \emph{philosophical pragmatism.}
Planning theorists seek to guide practitioners in what they can actually \textit{do} to resolve conflict and navigate value differences.
We argue that there are similar shifts in algorithmic fairness, analogous to the evolution of pragmatic thinking in the planning literature. 
To address algorithmic fairness problems \textit{in practice}, we believe that scholars can learn from the ways in which urban planners have navigated the messy and political reality of their work for many decades.

\begin{figure*}
\centering 
  \includegraphics[width=0.9\textwidth]{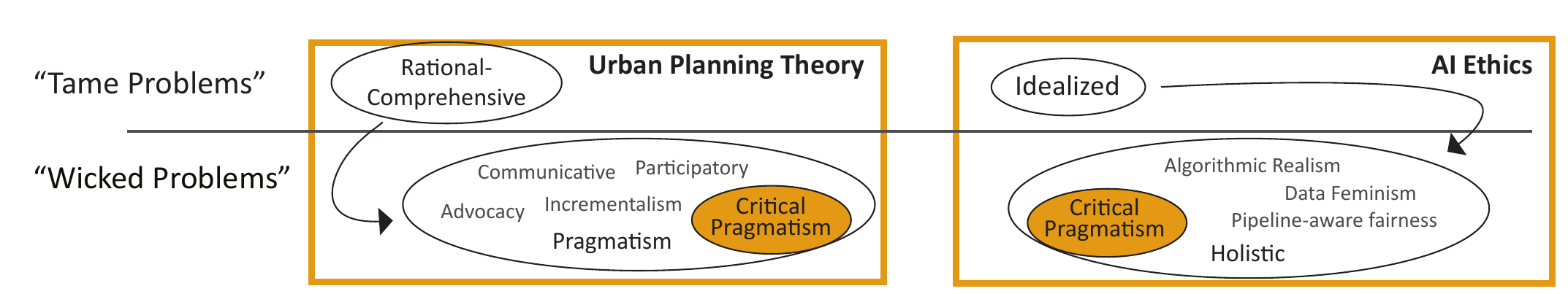}
  \caption{Connections are depicted between urban planning theory and AI ethics. Rational-comprehensive planning and idealized fairness definitions are more apt for ``tame'' or ``benign'' problems, as discussed in \citet{rittel_dilemmas_1973}, while pragmatic planning and holistic fairness address ``wicked problems.'' These terms are defined in Sections~\ref{fairness_planning},~\ref{wicked-problems-ml}, and~\ref{existing-alg-solutions}.}
  \label{fig:diagram}
\end{figure*}

Critical pragmatism attempts to resolve the practice-theory tensions that \citet{rittel_dilemmas_1973} first identified with ``wicked problems.'' It draws on prior theoretical work in urban planning and is grounded in empirical insights (e.g., first-hand accounts from urban planners). It addresses planning theory’s ``dirty little secret'' -- that even, despite extensive theorizing, ``somebody still has to do the work''~\citep{forester2017evolution}.

\textbf{Our primary contribution in this paper is to propose a set of four recommendations based on the urban planning theory approach of critical pragmatism.} We recommend that practitioners: (1) reflect-in-action,
(2) hold public meetings that involve deliberation, (3) listen critically and seek out creative negotiations, and (4) share practice-oriented stories.
We apply our four recommendations to three case studies: mortgage lending, school choice, and activism against feminicide.  In each case study, we describe the current algorithm and context.
We then discuss how our recommendations might improve on the current approach and where they fall short.

Our recommendations complement existing scholarship in algorithmic fairness and AI ethics, which has similarly called for increases in participation and critiqued power~\citep{suresh2022towards, delgado2023participatory, sloane2022participation,kulynych2020participatory, suresh2024participation, costanza2020design, d2023data}. Still, our recommendations formalize and advance important points that we believe are under-emphasized in the literature.

First, several of our recommendations are aimed at individuals, and can be utilized without institutional resources. These recommendations are useful for generalizing across the diverse and context-specific problems that practitioners often encounter.
Second, critical pragmatism emphasizes important social and dialogical tools needed to adequately address AI ethics concerns. Participation is useful not as a co-design tool to elicit information, but to advance a democratic and deliberative process, mediating across seemingly intractable value conflicts and building trust. This perspective builds on philosophers like J\"urgen Habermas and Paulo Freire~\citep{forester2017evolution}, whose work on communication and education is less commonly cited in  AI ethics literature.

\textbf{To underscore this point, we distinguish \emph{deliberation} from participation.} Deliberation requires active engagement from stakeholders (similar to the kinds of listening sessions, working groups, and public meetings that are common in urban planning, but less so in conventional approaches to algorithm design or ML). It is not that these technical processes never involve participation from stakeholders, but in many settings, there is little emphasis on communication -- \emph{listening} in particular -- as a tool for resolving conflict and trade-offs. 
As \citet{delgado2023participatory} note, many participatory ML projects consult with stakeholders only once, often through surveys or interviews. Sustained engagement involving genuine deliberation is relatively rare.
Still, in several instances where researchers have reflected on effective practices (e.g., in ML or AI for social good initiatives), such reflections have revealed the following elements of success: the sensitive manner in which information was communicated, the gradual building of trust and accountability through carefully structuring the participatory process~\citep{sendak2020human}, and ``real listening''~\citep{ai4si_food_rescue}. \textbf{We argue that these elements should be more clearly articulated as guidance for practitioners aiming to develop fair and ethical algorithmic tools.}

This paper is intentionally theoretical in scope. We do not claim that critical pragmatism will \emph{solve} all problems in AI ethics. In fact, the premise of our argument rests on the point that such problems are not solvable. 
Instead, by engaging with the urban planning literature, we offer tools that practitioners can use to incrementally forge a path forward.

The rest of the paper is structured as follows. In Section~\ref{fairness_planning}, we define urban planning and introduce the concept of wicked problems. 
We describe the evolution of pragmatic thinking in urban planning in Section~\ref{wicked-problems-ml}. Then, in Section~\ref{defining-critical-prag}, we introduce the framework of critical pragmatism, discuss foundational concepts, and note relevant critiques.
In Section~\ref{sec:applications}, we turn to recent work in algorithmic fairness and justify the analogy between the two fields. We briefly summarize different algorithmic fairness approaches, categorizing them as either idealized or holistic. We then discuss ongoing practical concerns with algorithmic fairness related to AI ethics and governance. In Section~\ref{call-to-action}, drawing on critical pragmatism, we provide four overarching recommendations that scholars and practitioners can flexibly use in their ongoing work. In Section~\ref{applying-pragmatism}, we apply these recommendations to several case studies. We conclude by encouraging scholars and practitioners in AI ethics to consider these recommendations drawn from urban planning theory.

\section{Defining Wicked Problems}
\label{fairness_planning}

Urban planning stems from the broader, interdisciplinary field of planning. Planning has been called the link between knowledge and action~\citep{friedmann1987planning}, the ``guidance of future action''~\citep{forester1999deliberative}, or the collection of a diverse set of practices~\citep{alexander2022planning}.
For many decades, scholars have debated different definitions~\citep{alexander1981if}. 
Planners navigate high-stakes political conflicts and guide the design, implementation, and impact of policies with evidence-backed decision-making and stakeholder engagement. The field is also closely related to architecture and design.  
Planning is: ``societal, future-orientated, non-routinised, deliberate, strategic and linked to
action''~\citep{alexander1981if}.

Planning also deals with ``wicked problems'' ---intractable and ill-formed problems that stand in contrast to the ``tame" or ``benign'' problems common in science and engineering~\citep{rittel_dilemmas_1973} (see Figure~\ref{fig:diagram}). Examples of wicked problems include serious policy issues with no straightforward solutions like poverty or climate change.

To provide a concrete example, consider housing affordability -- a concern in many cities around the world. Affordable housing is a central issue in urban planning, and one of the most common planning specializations according to the American Planning Association (\citeyear{apa_specializations}). In the United States (U.S.), there is currently no state or county where a renter who works full-time at minimum wage can afford a two-bedroom apartment~\citep{nlihc_out_of_reach}. Yet we know that housing access and stability affect life outcomes like employment, physical and mental well-being, educational achievement, and financial success \citep{desmond2016housing, gubits2015family}.

Housing affordability meets the 10-step criteria that \citet{rittel_dilemmas_1973} use to define wicked problems.
For illustrative purposes, we discuss the first one: ``there is no definitive formulation'' of the problem. The other steps are discussed in more detail in Appendix Section~\ref{sec:appendix_wicked_problems}.
This first step is true because there are different explanations for why housing is unaffordable. 
Some organizers and advocates call for more public housing \citep{cbpr_public_housing}. Others argue for housing vouchers~ \citep{cbpr_vouchers}. To some, zoning is the issue: upzoning neighborhoods should produce denser housing and reduce rents. 
Calls for upzoning are often met with resistance, due to fears about neighborhood change or displacement~\citep{brookings_commentary}. In land use decisions, affordable housing is often the site of intense conflict, with homeowners, renters, housing advocates, and government officials on very different sides of the issue.

\section{Urban Planning History Addressing Wicked Problems}
\label{wicked-problems-ml}

\citet{rittel_dilemmas_1973}'s ``wicked problems'' was a reaction to the emphasis on scientific and rational thinking in modern professions around the mid-20th century. This thinking produced the \textit{rational-comprehensive} model of planning \citep{whittemore_practitioners_2015} in which planners functioned much like scientific experts. 
Rational planning has been widely criticized with many real examples of failure in U.S. history \citep{lindblom2010science, schon_reflective_2017, scott1998seeing}.

Urban renewal stands out as one of the most emblematic and problematic examples. Shortly after World War II, the U.S. federal government sought to revitalize cities through redevelopment. At the time, rational-comprehensive planning dominated. Systems analysis, computational models, and quantitative analysis were the primary methods of urban planners \citep{hall1989turbulent}. Urban renewal ushered in a period of widespread slum clearance and new construction that left a legacy of destruction \citep{gans1959human, anderson1964federal, jacobs1992death}. As a result, local governments eliminated entire neighborhoods and displaced low-income residents of color \citep{hall1989turbulent, von2008lost, gans1959human}.

Urbanists frequently cite Jane Jacobs as one of the central figures opposing urban renewal \citep{scott1998seeing}. Jacobs saw the city as an ``immense laboratory of trial and error'' \citep{jacobs1992death}. She criticized mid-20th century planners like Le Corbusier whose ``City in the Park'' style of architecture became the blueprint for many U.S. housing projects~\citep{hunt2001went}
and Robert Moses who executed large-scale infrastructure projects in New York City (NYC), often at the expense of the residents living there. These planners had abstract, utopian visions of cities, divorced from social and historical context~\citep{callahan2014jane, jacobs1992death}. Jacobs condemned numerous examples of such planning failures: the decline of the Morningside Heights neighborhood in NYC following slum clearance and the development of a middle-income housing project, underutilized and run-down civic centers built according to the ideals of the modernist City Beautiful movement like the Benjamin Franklin Parkway in Philadelphia, or the homogeneity of single-use areas like Wilshire Boulevard in Los Angeles \citep{jacobs1992death}.

\subsection{The Ensuing Turn to Pragmatism}
\label{pragmatism}

The failures of rational planning are a warning to other fields. Rational planning had real and disastrous consequences on cities.
However, these failures also ushered in a shift toward pragmatism in urban planning (see Figure~\ref{fig:timeline}). Rittel and Webber's concept of ``wicked problems'' aligns with the onset of this shift. 
While influential across many fields, in urban planning their work has inspired newer generations of scholars to advance theory grounded in philosophical pragmatism \citep{forester2020five}.

Pragmatism is a philosophical tradition that argues against pre-determined \textit{a priori} beliefs and principles \citep{healey_pragmatic_2008}. Pragmatists prioritize practical experience. They believe that hypotheses about what works in the world should be continuously discovered and tested empirically \citep{healey_pragmatic_2008}. The rise of pragmatic thinking can be traced to three central figures: William James, Charles Sanders Peirce, and John Dewey \citep{bacon2012pragmatism}. Dewey in particular had a lasting influence on U.S. policy. His emphasis on experimentation, inquiry, and social justice influenced ideas on participatory democracy and popular support for social programs like the New Deal \citep{healey_pragmatic_2008}. \citet{deweypublicandproblems} writes:

\begin{quote}When we say that thinking and beliefs should be experimental, not absolutistic, we have then in mind a certain logic of method, not, primarily, the carrying on of experimentation like that of laboratories... Differences of opinion in the sense of differences of judgment as to the course which it is best to follow, the policy which it is best to try out, will still exist. But opinion in the sense of beliefs formed and held in the absence of evidence will be reduced in quantity and importance. No longer will views generated in view of special situations be frozen into absolute standards and masquerade as eternal truths. \end{quote}

Planning theorists have increasingly embraced pragmatic thinking--both older ideas from James, Peirce, and Dewey and more contemporary work from philosophers like J\"urgen Habermas, Richard Bernstein, and Richard Rorty \citep{healey_pragmatic_2008}. Examples of pragmatism in planning include: \citet{rittel_dilemmas_1973}'s critiques about viewing policy and planning professions as a science, \citet{lindblom2010science}'s work on incrementalism and experimentation, \citet{friedmann1973transactive}'s transactive planning approach to generate dialogue and navigate difference, \citet{davidoff1965advocacy}'s notion of plural plans and advocacy planning model, and \citet{forester_theory_2013}'s \emph{critical pragmatism}. We provide definitions for these early pragmatic approaches in Appendix Section~\ref{sec:app-pragmatic-planning-defn}. We focus on later approaches (reflective practice, deliberative practice, and critical pragmatism) in the following section.

\section{Defining Critical Pragmatism}
\label{defining-critical-prag}

Critical pragmatism is a unifying framework that builds on pragmatic planning approaches. Proposed by theorist John Forester, critical pragmatism is concerned with practice, not ideals. It focuses on real-world examples through interviews and case studies. 
It also grapples with deliberation and disagreement in public settings \citep{forester_theory_2013}. 

There are five key characteristics. 
First, it considers both process and outcomes, in part because of the ``co-constructed'' nature of how it approaches planning practice. Second, through its appreciation for different forms of knowledge, it encourages listening and interpretation. It considers the influence of power on relationships and claims about knowledge. Third, it is concerned with modes of communication and deliberation. Fourth, it offers lessons about process design and creative problem-solving. Fifth, it can help us shift ``from a deconstructive skepticism toward a reconstructive imagination'' \citep{forester_theory_2013}. Critical pragmatism does not force us to pessimistically accept reality as it is (as a simple ``pragmatic'' approach might), but remains optimistic that there are creative, alternative paths forward.

Many insights in critical pragmatism arise directly from analyzing stories about planners' actual experiences. Forester believes that planners learn through storytelling with friends and colleagues -- ``practice stories''~\citep{forester1999deliberative}. Drawing on the work of contemporary pragmatists like Habermas~(\citeyear{habermas}) and Freire's emphasis on dialogue and critical pedagogy~(\citeyear{pedagogy}), Forester connects pragmatic methods to learning in groups and democratic deliberation. He argues for an active and interpretive stance to communication--what he calls ``hermeneutic praxis''~\citep{forester_theory_2013, forester2017evolution}. This method involves understanding how meaning and information shape action \citep{forester1982planning} --- similar to an interpretive and dialogical approach to policy analysis \citep{wagenaar2014meaning}. 
\citet{forester_theory_2013} writes:

\begin{quote}The critical pragmatist...will attend not only to the consequences of framing a problem in a certain way, but to the contingencies of the relations of power and authority that can make alternative frames and knowledge claims more or less plausible in the first place.\end{quote}

\subsection{Reflective and Deliberative Practice}
\label{reflective-deliberative-practitioner}

Two concepts are foundational to critical pragmatism: reflective and deliberative practice. In this section, we define both separately. However, we will broadly refer to them together as the two concepts go hand in hand.

\subsubsection{Reflective Practice} Forester's critical pragmatism builds on Donald \citet{schon_reflective_2017}'s book \textit{The Reflective Practitioner}, which was first published in 1983. Sch\"on's concept of the ``reflective practitioner'' influenced many professional and design-oriented fields. This theory takes inspiration from earlier work in pragmatism, like Dewey's ideas about inquiry and trial-and-error. Sch\"{o}n's legacy was to contribute a new approach to the ``epistemology of practice''--rooted in what practitioners actually do and dissolving the sharp boundary between theory and practice. \citet{schon_reflective_2017} writes:

\begin{quote}When someone reflects-in-action, he becomes a researcher in the practice context. He is not dependent on the categories of established theory and technique, but constructs a new theory of the unique case... Because his experimenting is a kind of action, implementation is built into his inquiry. Thus reflection-in-action can proceed, even in situations of uncertainty or uniqueness..."
\end{quote}

\citet{schon_reflective_2017} describes a series of steps to effectively learn and generalize from the many context-specific situations that arise in practice: (1) try one approach to framing the problem; (2) test this framing and consider any challenges that arise; (3) revise the initial framing of the problem by reflecting on the results---a process known as ``backtalk'' (i.e., unexpected results). While each problem may be unique, through this approach, one can eventually develop a ``repertoire'' of similar experiences to inform problems in new situations~\citep{schon_reflective_2017}.

\subsubsection{Deliberative Practice} Forester finds Sch\"on's image of the individual practitioner engaged in reflective practice too limiting \citep{schon_reflective_2017, forester_theory_2013}. He expands on the notion of reflective practice by considering social and political context, multi-stakeholder settings, democratic and deliberative processes, and learning in groups. Thus, we can view Forester's deliberative practitioner as an evolution of Sch\"on's reflective practitioner.

\textbf{Deliberation is distinct from participation. Forester uses deliberation in the political sense, to denote group learning and dialogue}. This definition aligns with contemporary pragmatists who connect the pragmatic emphasis on inquiry and problem-solving with collective deliberation \citep{anderson2020pragmatist, misak2004making}.

Both \citet{schon_reflective_2017}'s notion of ``reflection-in-action" and \citet{forester_theory_2013}'s deliberative practitioner offer ways to tackle wicked problems. Critical pragmatism additionally incorporates ideas about how practitioners can navigate communication amidst conflict, value differences, and power.
We now discuss some critiques of critical pragmatism before turning back to a discussion of theoretical work in algorithmic fairness and comparing the two fields.

\subsection{Critiques and Limitations}
While Forester's work on critical pragmatism is included in many urban planning curricula, it has also received criticism, often for its overarching focus on communication. Communicative planning, exemplified by work from Patsy Healey~(\citeyear{healey_planning_1992}) and Judith Innes~(\citeyear{innes1995planning}), emphasizes consensus-building, collaboration, and dialogue. Critics of communicative planning raise several issues: (1) indifference to structural power dynamics (Forester's work is perhaps a counterexample~\citep{forester1980critical, forester1988planningpower}, though some might argue that he does not thoroughly consider structural dimensions of power); (2) the ambiguous definition of ``planners'' that assumes neutrality and benevolence~\citep{allmendinger2002communicative}; (3) the assumption that consensus is desirable~\citep{huxley2000limits}; and (4) the dominance of Western ideals and theories, which casts planning as both universal and unproblematic~\citep{huxley2000new}. Many of these critiques overlap with broader concerns about the role of planning in upholding state power, and relate to criticism of more foundational theorists, such as Habermas~\citep{huxley2000new}.

We encourage readers to consider these critiques and use them to frame similar discussions in AI ethics. 
Critical pragmatism advances a set of methods that can provide practical, incremental~\citep{lindblom2010science} guidance about how to move forward. However, we do not suggest that critical pragmatism is a catch-all solution to any AI ethics problem, or even broader societal issues. Alternative approaches (e.g., grounded in political economy and critical theory) may be more useful in some settings~\citep{huxley2000new, yiftachel2000debating}.

\section{Connections between Urban Planning Theory and AI Ethics}
\label{sec:applications}

As algorithmic systems influence many high-stakes, societal decisions, the framing of wicked problems has become increasingly relevant to the field of AI ethics.
Algorithms often must confront the underlying social and political dynamics that make such decisions inherently difficult, even in non-AI settings. Recent examples include: predictive policing~\citep{lum2016predict, green2019smart}, AI hiring~\citep{fabris2025fairness, robinson2022voices, gerchick2025auditing}, and school choice algorithms~\citep{schoolchoice_nyc_markup2, schoolchoice_nyc_markup}. Competing definitions of fairness and equity further complicate these decisions. To illustrate the connection to ``wicked problems,'' we provide a worked example in the appendix related to mortgage lending algorithms.

The overlap between urban planning and AI ethics -- both grapple with ``wicked problems'' -- suggests there may be value in learning from urban planning theory.
Though prior work in AI ethics has cited \citet{rittel_dilemmas_1973},\footnote{See prior work like~\citet{dobbe2021hard, scantamburlo2021non, strauss2021deep, green2019smart}.} the discussion of ``wicked problems'' in the literature has been relatively brief and the origins of the term within urban planning remain unexplored. The connection between the two fields is especially relevant because, as we discuss in the next section, the evolution of theoretical scholarship in AI ethics over the last decade mirrors that of urban planning.

\subsection{The Comparison between Fields}
We recognize that not all algorithmic fairness problems may compare to urban planning, but there are several reasons why the comparison between urban planning and algorithmic fairness is justified and even useful. First, as a scoping decision, we primarily focus on allocative harms in algorithmic fairness, as opposed to representational harms~\citep{barocas-hardt-narayanan}. Allocation decisions are widespread in both fields.

Second, urban planning is often discussed in the context of local (city or neighborhood-level) problems. 
This perspective overlooks the fact that urban planning involves local responses to broader societal issues, much like how data scientists and engineers may need to adapt scalable technologies to specific contexts.
While \emph{urban} planning deals with small-scale problems, the field of planning encompasses regional, economic, and environmental planning, which can all occur at non-local scales and involve extensive coordination (e.g., a national program). Algorithmic fairness problems similarly can vary in scale. We believe our recommendations extend even to settings involving large-scale technology. Participatory processes can reasonably occur even at the national level (e.g., see \textit{Voices in the Code}~\citep{robinson2022voices}). Decisions about which stakeholders to include, communication choices, and problem framing may be even more critical and high stakes in such settings.

Third, many view urban planning as a field involving public-sector stakeholders, who may already approach ``wicked problems'' with a normative frame of justice and equity.
While many urban planners do work in the public sector, urban planning encompasses a diverse set of roles in both public (e.g., local government) and private organizations (e.g., development companies), similar to roles in data science and engineering. We believe our recommendations, especially those aimed at individuals, are relevant, irrespective of a specific industry or role.

Lastly, we acknowledge that our recommendations may not be appropriate in some AI contexts where priorities are more narrowly tied to specific, private business interests. Our recommendations may be most effective when there is at least some commitment to developing fairer algorithms and resolving conflict, but disagreement about how. Additionally, even privately developed AI tools can have broader societal impact that warrants regulation. NYC Local Law 144~\citep{locallaw144} is one example.

\subsection{Algorithmic Fairness Approaches Mirror Urban Planning History}
\label{existing-alg-solutions}

Borrowing
terminology from \citet{barocas-hardt-narayanan, fazelpour_algorithmic_2020}, we classify two distinct approaches to fairness  that have emerged in the literature on AI ethics: idealized vs. holistic fairness.
This classification captures ongoing tensions in the field (see Figure~\ref{fig:diagram}): the limitations of idealized abstractions in comparison to the broader contextual and philosophical understandings of fairness \citep{green2022escaping}. \textbf{The difficulty scholars face in reconciling different fairness definitions parallels the same dilemma Rittel and Webber confronted when proposing wicked problems.}

\subsubsection{\textbf{Idealized Fairness}}
Idealized fairness definitions often formalize different fairness criteria mathematically and make simplifying assumptions that may be unrealistic. These approaches are precise and tractable, but may be too narrow. In real-world settings, they can lead to unfair outcomes \citep{green2020algorithmic, green2020false, kalluri2020don, chohlas2023designing}. For example, as researchers have shown, satisfying fairness via multiple formal definitions in many real-world instances can even be \textit{impossible} both theoretically and in practice \citep{chouldechovadisparateimpact, kleinberg2016inherent}. 
Many established fairness frameworks and toolkits tend to use these methods (e.g., the \emph{fairlearn} library ~\citep{weerts2023fairlearn}).

\subsubsection{\textbf{Holistic Fairness}}
In reaction to the limitations of idealized fairness approaches, researchers have turned to more holistic frameworks, drawing on scholarship from science and technology studies (STS), law, philosophy, and economics \citep{pessach2023algorithmic, jacobs2021measurement, binns2018fairness, heidari2019moral, arif2022towards, lee2021formalising, selbst2019fairness, davis2021algorithmic}. 
We classify approaches as holistic if they equate fairness with more expansive notions of equality and justice.

Guidance on implementing holistic fairness in practice, however, is not clear. 
Many holistic approaches treat fairness as context-dependent or as a social and political construct.
These methods still require value judgments and lack a principled process. For example, operationalizing equal opportunity doctrines from political philosophy \emph{in practice} is still far from straightforward~\citep{arif2022towards, heidari2019moral}. Likewise, \citet{selbst2019fairness}'s work drawn from STS theory argues for prioritization of social context and process over outcomes, which requires thoughtful and subjective judgment.

\subsubsection{The ``Participatory Turn'' in Algorithmic Fairness}
\begin{figure*}[!htbp]
\centering
        \centering
        \includegraphics[width=\linewidth]{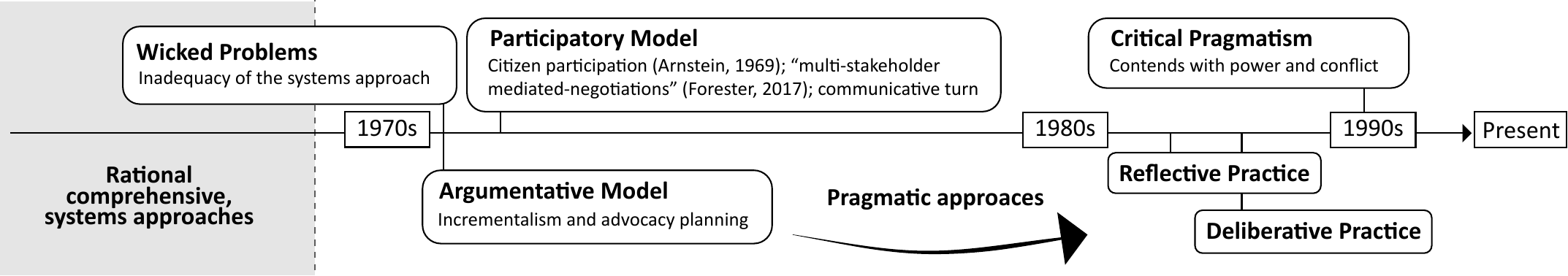}
  \caption{The influence of pragmatic ideas on planning theory since \citet{rittel_dilemmas_1973}'s ``wicked problems'' (represented by the arrow). Based on \citet{hoch1984pragmatism, healey_pragmatic_2008,forester2017evolution}.  
  While AI ethics has analogously taken a ``participatory turn''~\citep{delgado2023participatory}, reflective and deliberative practice and critical pragmatism have yet to be applied widely.}
  \label{fig:timeline}
\end{figure*}

In recent years, interest has grown in applying participatory methods to ML and algorithm design --- paralleling a similar historical development in urban planning~\citep{delgado2023participatory} (see Figure~\ref{fig:timeline}). Prior work has advanced the idea that engaging stakeholders may lead to more equitable algorithmic systems, shift power, and mitigate harm \citep{suresh2022towards, birhane2022power, kulynych2020participatory, costanza2020design, d2023data, koenecke2023popular}. Recent examples include a case study of QueerInAI (\citeyear{queerinai2023queer}), a community-led, grassroots organization, 
 and a participatory algorithm design framework applied to on-demand food donation \citep{lee2019webuildai}. Several other examples are documented in~\citet{Kim_Balloli_Karimian_Bondi-Kelly_Fish_2025}.

\subsubsection{A Pragmatic Turn in Algorithmic Fairness?}
Recently, there has also been a shift toward pragmatic thinking, such as \citet{green2020algorithmic}'s work on ``algorithmic realism.'' Drawing on the influence of ``legal realism'' on American legal thinking, they argue that designing fair algorithms must be political, porous, and contextual. Recent calls for a pipeline-aware approach to fairness \citep{black_toward_2023, akpinar2022sandbox, suresh2021framework} can also be viewed from a pragmatic perspective; they advocate for an iterative process and are concerned with how bias arises in practice. Lastly, \citet{watson2024competing} propose ``sociotechnical pragmatism'' as a compromise to resolve tensions between AI dogmatism and skepticism.

\subsection{The Limits of Current Fairness Approaches in Practice}
\label{gap-theory-practice}

The shift toward participatory and pragmatic methods reflects a growing recognition that current approaches to algorithmic fairness often break down in practice. \citet{holstein2019improving} note that practitioners' needs are at odds with the priorities of most fairness research. The authors call attention to the fact that while much of the algorithmic fairness literature focuses on models, practitioners think primarily about data collection. 
\citet{deng2022exploring} and \citet{lee2021landscape} both explore how industry practitioners use fairness toolkits, observing a need for more contextualization, integration with existing workflows, and accessible communication with non-technical audiences.
 
There are many more challenges that practitioners face \citep{veale2018fairness,veale2017fairer, madaio2020co,madaio_assessing_2022, law2020designing, richardson2021towards, gosciak_hvi}, even when considering holistic fairness approaches. Current fairness efforts are often ad-hoc and limited by organizational constraints~\citep{yan2025fairness}. Some papers offer practical design suggestions: semi-automated bias detection tools with practitioner input, fair ``checklists''
~\citep{madaio2020co, madaio2024tinker}, or collaborative, experiential knowledge bases~\citep{veale2017fairer}. The need for these solutions further illustrates the tension between theory and practice.

Furthermore, concerns related to the \emph{mis}-use of participatory methods have arisen. Scholars have raised well-founded critiques that participation in ML and AI is largely ``consultative,'' narrow, and indifferent to power dynamics \citep{delgado2023participatory, feffer2023preference, birhane2022power, sloane2022participation, corbett2023power, groves2023going}.
Compounding this criticism is the fact that most recent work documenting participatory processes focuses on context-specific, in-depth case studies, from which it can be hard to generalize~\citep{suresh2024participation}. 
There is little cohesive and practical guidance about how to structure participatory processes, particularly amidst deep-seated value conflicts. While many note the importance of participation and involving stakeholders, a greater emphasis on listening, deliberation, and modes of communication can make existing participatory approaches more effective.

We turn to urban planning to fill the gap in pragmatic approaches to wicked problems in algorithmic fairness.
Urban planning is a field that has championed participation and communication.
Drawing on critical pragmatism, we develop flexible recommendations that data science practitioners and organizations can apply to their ongoing work.

\section{Urban Planning Tools for AI Ethics}
\label{call-to-action}

The urban planning shift away from rational-comprehensive planning toward pragmatism, as described in Section~\ref{pragmatism}, reflects the tensions that arise between idealized and holistic fairness definitions.
In Section~\ref{existing-alg-solutions}, we discussed how prior work in AI ethics has begun to advocate for participatory and pragmatic approaches, but in Section~\ref{gap-theory-practice}, we noted the shortcomings of these methods and ongoing challenges that practitioners face. 
We can apply learnings from urban planning in addressing similar shortcomings: specifically, regarding pragmatic ideas---and critical pragmatism in particular---as described in Sections~\ref{pragmatism} and~\ref{defining-critical-prag}.

Critical pragmatism provides a set of problem-solving methods that help practitioners answer the question: What can we actually \textit{do}? Here, drawn from the core tenets of critical pragmatism, we propose four recommendations.

\begin{enumerate}
    \itemsep0em
    \item \textbf{Reflect-in-Action:}
    Practitioners should approach algorithmic fairness like improvisation, iteratively and creatively. This is the essence of the pragmatic tradition. Instead of seeking out fixed ``normative truths" about what constitutes ethical and fair algorithms, practitioners should remain open to ``intelligently updating'' their beliefs as they experiment with different approaches \citep{anderson2020pragmatist}. 
    Capturing this dynamic, \citet{schon_reflective_2017}'s reflection-in-action is a useful heuristic. It offers a way to manage the context-dependent nature of many algorithmic fairness problems, by approaching the set of unique problems as a professional ``repertoire''~\citep{ schon_reflective_2017}. Individuals can reflect-in-action, even when there is no institutional support.
    Notably, while pragmatism rejects the rigidity of formal models that function as absolute truths, it appreciates formalisms (e.g., equations, mathematical models) as useful for experimentation and inquiry.
    \item \textbf{Practice deliberation through public meetings:} Urban planning is known for long public meetings and community engagement processes, but this kind of deliberation -- mediating conversations among divided groups -- is central for resolving conflict. In a similar vein, participatory methods are becoming more common in ML and AI, but these methods are not necessarily deliberative nor valued as such. Scholars in both planning and AI ethics have cautioned against narrow and extractive forms of participation~\citep{arnstein1969ladder, monno2012tokenism, sloane2022participation, delgado2023participatory, corbett2023power}. Even more justice-oriented approaches often resort to interviews as opposed to sustained public meetings involving real dialogue.  We argue that deliberation is not just important for AI governance~\citep{buhmann2023deep}, but for the entire AI development process --- from framing the problem appropriately to safely deploying a tool. \citet{forester1999deliberative}'s \textit{The Deliberative Practitioner} offers practical guidance about how to structure deliberative processes --- e.g., framing public meetings as discussions rather than debates, and encouraging informal participatory rituals that build trust \citep{forester1999deliberative}. 
    These meetings are not meant to function like one-sided presentations. They are more similar to community town halls or other political spaces that encourage active participation along with bi-directional communication and learning.
    They can involve large public discussions or small groups, hundreds of regular meetings or only one.
    \item \textbf{Listen critically and seek out creative negotiations:} Attend to the role of communication in building trust and resolving conflict.
    This requires an active listening role, what Forester calls ``hermeneutic praxis''~\citep{forester_theory_2013} --- a critical interpretive stance to communication that involves understanding both what is said and not said (e.g., the meaning, emotion, and connotation hidden within specific word choices). Critical pragmatists would question stakeholders' claims to legitimacy --- e.g., claims of expertise as a ``professional'' or over-confident promises with statements like ``trust me''~\citep{forester_theory_2013}. They would consider how power and status may affect communication: bias a story, distort information, or advance a specific viewpoint~\citep{forester1982planning}. These skills are not often valued in technical fields and are more common in qualitative research~\citep{josselson2004hermeneutics}.
    Creativity is also important, as planners seek to resolve conflict in seemingly impossible settings. One should not ``presume impossibility,'' and instead attempt to shift power dynamics by anticipating information distortions or through mediation~\citep{forester1982planning, forester_theory_2013}.
     \item \textbf{Share practice-oriented stories:} \citet{forester1999deliberative} argues that practice-oriented storytelling among friends and colleagues, is an important part of learning. These stories
     are not trivial, nor are they traditional interviews or case studies. They are lengthy, minimally edited, first-hand accounts directly from practitioners about their experiences. They are ``rich, morally thick, politically engaged, and organizationally practical'' \citep{forester1999deliberative}. Much of Forester's own work is based on insights derived from practice stories. In fact, understanding his theoretical contributions is most useful in the context of specific examples, which he analyzes over several books \citep{forester1999deliberative,forester1988planningpower, forester1993criticaltheory}. 
     Forester provides a starting point for collecting these stories – e.g., asking ``How'' rather than ``Why'' questions or  ``actor-focused'' rather than ``spectator-focused'' questions \citep{forester2012learning}.
     Practice stories view practitioners as social actors, confronting the messy reality of their work \citep{wagenaar2011beckon}.
\end{enumerate}

Our recommendations touch on existing calls to action in AI ethics: increase participation, critique power structures, and account for context~\citep{suresh2024participation, costanza2020design, d2023data, koenecke2023popular, Kim_Kuehnert_Lazar_Singh_Heidari_2025}. Going a step further, our recommendations are aimed at operationalizing and generalizing from these efforts. Beyond simply consulting stakeholders, we emphasize dialogue, listening, and concern for different modes of communication. These are critical considerations in urban planning that have not been fully utilized in AI ethics nor formalized as specific calls to action. Concepts like reflection-in-action and practice-oriented stories complement these recommendations by helping practitioners learn from and generalize across varied, context-specific tasks.

\section{Applying Critical Pragmatism to Algorithmic Fairness: Case Studies}
\label{applying-pragmatism}

In this section, we envision what a critical pragmatism approach for AI ethics might look like in three case studies. 
For each case study, we discuss the current approach and how critical pragmatism might lead to improvements.
We also identify instances where the framing of critical pragmatism makes explicit strategies that worked well and where critical pragmatism might fall short. We further connect our recommendations to specific points in the algorithm development pipeline in Appendix Section~\ref{sec:guidance}. We believe our recommendations overall can help with generalizing from these case studies to other settings.

Context varies across each of our case studies. In the first case study, ethical concerns have led to algorithm changes only recently. In the second case study, the algorithm was motivated by concerns related to equity and efficiency. The third case study is grounded in data feminism, which embeds a wide range of ethical concerns holistically in the project.
These three case studies also represent differences in geographic scale. We believe our recommendations apply regardless of scale, whether national or local.

\subsection{Case Study 1: Mortgage Lending}
\label{critical-pragmatism-example-ml}
\textit{The Problem.}~
Given the history of housing inequality in the U.S., fairness and discrimination are paramount concerns for the government regarding mortgages. 
The Equal Credit Opportunity Act (ECOA) and the Fair Housing Act explicitly forbid discrimination in lending against many sensitive attributes (such as race, gender, national origin), and through the Home Mortgage Disclosure Act, financial institutions are required to disclose data on lending decisions to ensure the ECOA has been upheld~\citep{akinwumi2021ai}.
However, even though explicit lending discrimination is rare in the present-day, prior work has shown that mortgage lending decisions and access to credit tend to be racially skewed~ \citep{markup_mortgage, mehrotra2024evidence}. 

\textit{Current approach}.~The use of algorithms in mortgage lending decisions is not new. A 2019 report from \textit{The Markup} demonstrated how opaque automated underwriting practices and outdated credit scoring algorithms like ``Classic FICO'' have led to racially biased outcomes~\citep{markup_mortgage}. Furthermore, they show how racially disparate outcomes persist even when controlling for many kinds of public data that could potentially explain such disparities.
While technical methods have shown promise at reducing inequities, even with biased historical data~\citep{merrill2024improving, singh2022developing, lee2021algorithmic, nguyen_shelterforce, coston2021characterizing}, such methods are largely irrelevant, as a key impediment to updating the algorithm is political.
Until recently, federal agencies have resisted updating ``Classic FICO,'' even though simple changes (e.g., including rental payments or cashflow in credit-scoring models) could increase credit access to more borrowers~\citep{markup_mortgage, wsj_fico, kumar2022equalizing}.

\textit{Context.}~
Relevant stakeholders include: U.S. federal government agencies, financial institutions, and nonprofit advocates.
Financial institutions aim to reduce risk, nonprofit advocates advance equity concerns, and the federal government aims to serve the public interest.
While private companies and nonprofits can develop fairer models on their own~\citep{nguyen_shelterforce}, many lenders may still opt to follow federal policy guidelines to ensure funding guarantees from the government. Technical fairness methods, including recent proposals from academics, may have little effect without concrete changes in federal policy. While large-scale public engagement processes may be challenging to coordinate at the national level, the federal government is best positioned to oversee such processes and can make far-reaching policy changes.

\textit{Alignment with Holistic Fairness}: Low. Current practices in mortgage lending, described above, do not reflect a critical pragmatism approach. When considering academic contributions and technical mitigation strategies, most fairness interventions are focused on improving prediction outcomes and evaluation metrics --- not the process of developing the tools themselves. Many of these interventions have been primarily speculative and theoretical.

\textit{Applying critical pragmatism:} Critical pragmatism would involve a more political and deliberative process through public meetings.
The federal government's automated underwriting process, which is both opaque and largely one-sided, has not supported this kind of process historically~\citep{markup_mortgage}.
Government agencies like the Consumer Financial Protection Bureau (CFPB) and Federal Housing Finance Agency (FHFA) do solicit public feedback, e.g.: public comment on any newly proposed rule, requests for information, and listening sessions~\citep{cfpb_comment, fhfa_listening_sessions}. For written comments, the government evaluates them and decides whether to modify, withdraw, or proceed with each respective rule \citep{regulations_gov}. Requesting public comments results in detached, transactional interactions, not the collaborative and transformational deliberation of critical pragmatism where the actual process of negotiation and communication is important.

The FHFA's ``stakeholder forums and listening sessions" are a promising direction, but these largely include testimony from established industry representatives~\citep{fhfa_input, fhfa_public_engagement, fhfa_listening_sessions}. 
A critical pragmatism approach would seek to mitigate disparities in participation (reducing material or psychological barriers to participation) and would consider how power may shape  communication. 
By building trust gradually through open discussion, the hope is that a shared set of priorities may emerge. Adopting this kind of approach might have led stakeholders to accept simple methods to improve fairness sooner (e.g., including non-traditional inputs to credit scores~\citep{kumar2022equalizing, wsj_fico}).

\subsection{Case Study 2: School Choice}
\textit{The Problem.} School choice decisions in many urban areas are complex. NYC -- the largest city in the U.S. -- assigns students to middle and high-schools using the deferred acceptance (DA) algorithm. The algorithm matches students to schools using students' preferences and by ranking students based on academic performance, location, and whether the student receives free or reduced-price lunch. This algorithm is widely credited with improving the high school assignment process -- both reducing the number of unmatched students and increasing the number of students receiving their first-choice match~\citep{chalkbeat_da}.
Prior to 2003, the matching system was inefficient; almost a third of all students did not receive placements to schools they had initially selected~\citep{abdulkadirouglu2005new}.

\textit{Current approach.} The DA algorithm has led to significant improvements.
The new system is more transparent and the DA algorithm is strategy-proof, meaning it is not possible for students to strategically manipulate outcomes. 
Still, students may feel unfairly matched, especially if they do not understand the DA algorithm or fail to match to their top schools. 
News reports highlight the system's informational complexity and the burden on parents, which may impact students with fewer resources and more limited support systems~\citep{chalkbeat_da}.
To counteract these effects, the city has explored informational and personalized interventions, which empirical evidence suggests are successful~\citep{cohodes2025informational, peng2025undermatching}.
But, NYC's school matching process continues to reinforce existing socioeconomic and racial inequalities, particularly through selective admissions screens~\citep{chalkbeat_unequal}.

A similar problem occurred in San Francisco where the city voted to abandon a school choice algorithm after finding that it exacerbated inequality. \citet{robertson2021modeling} examined the design and implementation of the algorithm using a Value Sensitive Design approach. They observed several misalignments between modeling assumptions in the algorithm and real-world challenges. Many of these -- such as inequitable gaps in access to information or issues of trust -- could be mitigated if the design of the algorithm had adopted a critical pragmatism approach.

\textit{Context.} This case study primarily involves public-sector employees and parents. While public-sector employees may care about broader equity concerns (which motivated the adoption of the DA algorithm in the first place), parents are likely more invested in their children's welfare. An effective critical pragmatism approach would recognize these sometimes-competing priorities and seek resolution.
The fact that the school district manages the DA algorithm provides legitimacy and support for carrying out several of our recommendations, such as organizing public meetings. The local scale can also facilitate deeper engagement with stakeholders, though notably NYC's school district serves roughly 900,000 students~\citep{nycps_data}.

\textit{Alignment with Holistic Fairness}: Moderate.  A strength of school choice algorithms like DA is the ability to easily simulate outcomes under a range of conditions. Simulation, much like the sketches that \citet{forester1999deliberative} discusses, can be a tool for the iterative and experimental reflection-in-action process. Reflection-in-action is a method that practitioners can use in any setting, when approaching a new problem or task. A weakness of current approaches to school choice -- exemplified in both NYC and San Francisco -- is the lack of any deliberative or in-depth public engagement process.
One of the main limitations that researchers have cited in prior work is a mismatch in information: misinformed perceptions from parents about how school choice algorithms work and inadequate guidance for students~\citep{cohodes2025informational}.

\textit{Applying critical pragmatism:} Deliberation and public meetings could address this problem. These might include interactive workshops and open listening sessions where those responsible for designing and implementing the algorithm are forced to confront the nuanced and conflicting values of different stakeholders.
The format of these meetings is critical and importantly helps to develop trust, which has been cited as a significant issue in San Francisco~\citep{robertson2021modeling, robertson2022not}.

Critical pragmatism complements existing methods like value sensitive design~\citep{zhu2018value}. Importantly, a critical pragmatism perspective would consider how power and resource differences shape participation (e.g., distorting information, eroding trust, or influencing the opinions stakeholders publicly express~\citep{forester1988planningpower}). In response, critical pragmatists would adopt an interpretive, ``hermeneutic'' stance to communication. This stance can help with the broader goal of resolving deep-seated value conflicts, particularly when there are pronounced power imbalances and issues of trust.  For example, in the school choice setting, differences in how parents perceive fairness is intertwined with race and socioeconomic status~\citep{nguyen2024definitions}.

A real example of a more deliberative process is an urban planning initiative for NYC School District 15 (D15). The D15 Diversity Plan has been an effort to integrate one of the city's most socioeconomically and racially segregated school districts prior to 2017~\citep{wxy_d15}. It built on years of advocacy, established a multi-stakeholder working group, and convened interactive workshops for families and educators -- providing childcare, food, transit passes, and translation to reduce barriers for attending~\citep{wxy_d15}. The D15 Diversity plan resulted in the removal of all selective screens and a choice-based lottery system with priority for low-income students. As of the 2022-2023 academic school year, District 15 has transformed from the second most segregated school district in the city to the 19th (out of 32)~\citep{wxy_d15_article, wxy_d15_report_followup}.

\subsection{Case Study 3: Feminicide Counterdata Collection}
\textit{The Problem.} We now turn to counterdata collection for monitoring feminicide. Counterdata is a collective, activist response to produce data in settings where powerful institutions (e.g., governments or corporations) fail to do so~\citep{d2023data}.
A collaboration across both activist and academic organizations, this project used intersectional feminist principles and participatory methodologies to co-design an ML system that identifies relevant news reports of feminicide for activists. The project started through interviews with representatives from both activist and civil society organizations~\citep{d2022feminicide}. Through these interviews, researchers identified the potential for ML to reduce the burden on activists, who regularly sort through violent and upsetting content, by partially automating the detection of feminicide cases in news articles~\citep{d2022feminicide, suresh2022towards}.

\textit{Current approach.}
The feminicide counterdata collection project was developed in close alignment with \citet{d2023data}'s work on data feminism. As a result, it demonstrates many of the same values that underlie a critical pragmatism approach. For example, challenging power is one of the core principles of data feminism, and central to this project. The researchers specifically worked with activist groups, rather than well-resourced government agencies or NGOs. They sought to develop a model with high performance on intersectional feminicide cases, not only cases for majority groups~\citep{suresh2022towards}. The project also built on participatory methodologies through a collaborative co-design process that viewed ``participation as justice''~\citep{suresh2022towards, sloane2022participation}.

\textit{Context.}~
This case study involves activists and academics, who appear relatively aligned in values and views. This dynamic could be the product of the participatory co-design approach and also the intentional decision to prioritize partnering with activists, as opposed to other institutional actors. Though the project does involve participation from stakeholders, the emphasis on trust and deliberation may not be as crucial as in other case studies because there are few high-stakes instances of conflict. The project operates with few regulatory constraints, and the resulting output is closely aligned with the priorities of the activist partners.

\textit{Alignment with Holistic Fairness}: High. This project reflects high alignment with holistic fairness approaches, as it is based on data feminism~\citep{d2023data}. As \citet{suresh2022towards} discuss, the co-design ML process was iterative and collaborative, involving stakeholder input through regular evaluation~\citep{d2022co}. It was also sensitive to context and critical of power. Given this background, critical pragmatism helps clarify what worked well.

\textit{Applying critical pragmatism:} In particular, this project exemplifies the value in our third recommendation, ``creative negotiations''~\citep{forester_theory_2013}. Such negotiations are only possible through carefully listening to stakeholders. Given both the desire to memorialize and honor the individuals in the data, and to avoid rigid definitions of feminicide, the research team could have decided initially to forego using ML.  However, by listening to feedback from activists, they negotiated a more creative, alternative approach involving partial automation: an ML system that generates email alerts to reduce the emotional burden on activists, while activists still retain control of defining the relevance of specific feminicide cases. This negotiated outcome was the result of listening to stakeholders, interpreting the underlying meaning in what they shared, and using this interpretation to creatively identify points of overlap.

The feminicide counterdata collection project is also useful in demonstrating the trade-offs necessarily arising from a critical pragmatism approach and how it may fall short in some contexts. The research team in this case prioritized working with activists and civil society organizations as a more radical stance in challenging power structures whereas critical pragmatism might have taken a more moderate approach through compromise and negotiation (e.g., engaging with news organizations or government directly).

\section{Conclusion}
\label{conclusion}
In this paper, we characterize algorithmic fairness problems as ``wicked problems,'' which are typically the domain of policymakers and urban planners. This characterization suggests a broader analogy between the two fields. We argue that urban planning theory, specifically critical pragmatism, is useful to the AI ethics community.
Critical pragmatism focuses on what works \textit{in practice}, addressing gaps in the AI ethics literature between theory and practice. It is grounded in real stories from urban planners. Drawing on critical pragmatism, we offer four recommendations: (1) reflect-in-action, (2) practice deliberation through public meetings, (3) listen critically and negotiate creatively, and (4) share practice-oriented stories. These recommendations underscore the importance of skills often not emphasized in technical, data science settings, but which are crucial for resolving fairness and ethics concerns -- skills related to reflexivity, listening, and dialogue.

\section*{Acknowledgments}
We are grateful to numerous individuals who have provided invaluable feedback and support over the past few years. First, we appreciate John Forester for engaging with our interdisciplinary attempts to bridge two distinct fields, and who consistently replied to our many questions about critical pragmatism. Second, we are grateful to colleagues and faculty at Cornell, who reviewed this work and provided comments -- including Angelina Wang, Anthony Townsend, Emma Harvey, Gabriel Agostini, Isabel Corpus, Jae June Lee, Matt Franchi, and Travis Lloyd.

\bibliography{aaai2026}

\appendix
\section{Pragmatic Planning Definitions}
\label{sec:app-pragmatic-planning-defn}

We now define several different urban planning models (visualized in Figure~\ref{fig:timeline}). These terms come from \citet{forester2020five}'s work on different ``generations'' of theory that are tied to the pragmatic tradition of resolving practice-theory tensions.

\begin{enumerate}
    \item \textbf{Rational-comprehensive model:} This is the model we describe in Section~\ref{wicked-problems-ml}, and was the first theoretical approach to urban planning that Rittel and Webber argued against in their seminal work on wicked problems. It prioritizes rationality and scientific thinking. Typically, it involves identifying plan objectives and evaluating a set of alternative plans in relation to these objectives. A case study of rational planning appears in~\citet{black1990chicago}.
    \item \textbf{Argumentative model:} This is the second generation of planning, according to \citet{forester2020five}. Some of these ideas appear in Rittel and Webber's work~(\citeyear{rittel_dilemmas_1973}). The argumentative model solves problems by increasing the number of perspectives and ideas --- shifting away from the notion of the planner as an individual expert to several planners working together collaboratively. Both \citet{lindblom2010science}'s incrementalism and \citet{davidoff1965advocacy}'s advocacy planning approaches fall under this model. The incrementalist approach views policymaking like trial and error, a set of incremental steps that might lead a decision-maker to eventually craft the right approach. In contrast, advocacy planning has an emphasis on equity and social justice. \citet{davidoff1965advocacy} argued that urban planners need to support all stakeholder groups -- including those with fewer resources -- to produce ``plural plans'' that can be evaluated alongside plans from established institutions.
    \item \textbf{Participatory model:} As planners increasingly sought to meaningfully include different stakeholders alongside trained experts, the participatory model of urban planning gained traction. This approach to planning was motivated by foundational work like \citet{arnstein1969ladder}'s ``A ladder of citizen participation.'' Arnstein provided a typology of citizen participation to describe whether participatory processes meaningfully redistribute power. It ranges from ``consultative'' (almost no involvement) to ``citizen control'' (citizens retain all decision-making authority).
    Later work from theorists like Lawrence Susskind and Judith Innes focused more on mediation of conflict through ``mediated-negotiations'' \citep{forester2020five}. Their work emphasized the power of producing legitimate, mutually beneficial solutions~\citep{susskind1987breaking, innes2018planning}. \citet{arnstein1969ladder}'s ``ladder of citizen participation'' has influenced several papers in AI ethics such as~\citet{tseng2024data, sloane2022participation, delgado2023participatory}. However, in prior work on participation, there has been much less of a focus on the role of mediation in resolving value conflicts and on the structure of deliberative processes that encourage genuine participation and build trust (what Forester envisions in \textit{The Deliberative Practitioner}~(\citeyear{forester1999deliberative})). 
    As \citet{garg2025heterogeneous} argues, participation in many real-world choice settings tends to be heterogeneous, with higher levels of participation often correlated with higher socioeconomic status and education levels. While it is important to design technical approaches that account for heterogeneous participation, there are many non-algorithmic approaches to increase participation (e.g., simple design changes or financial support to offset the costs of participation). Similarly, improving the methods of communication and reducing the barriers for engaging in deliberative processes is an important part of how urban planning approaches participation.
\end{enumerate}

\section{Recommendations \& Practical Guidance}
\label{sec:guidance}
\begin{figure*}[htbp!]
        \centering
          \includegraphics[trim={0 0.8cm 0 0.5cm},clip, width=\textwidth]{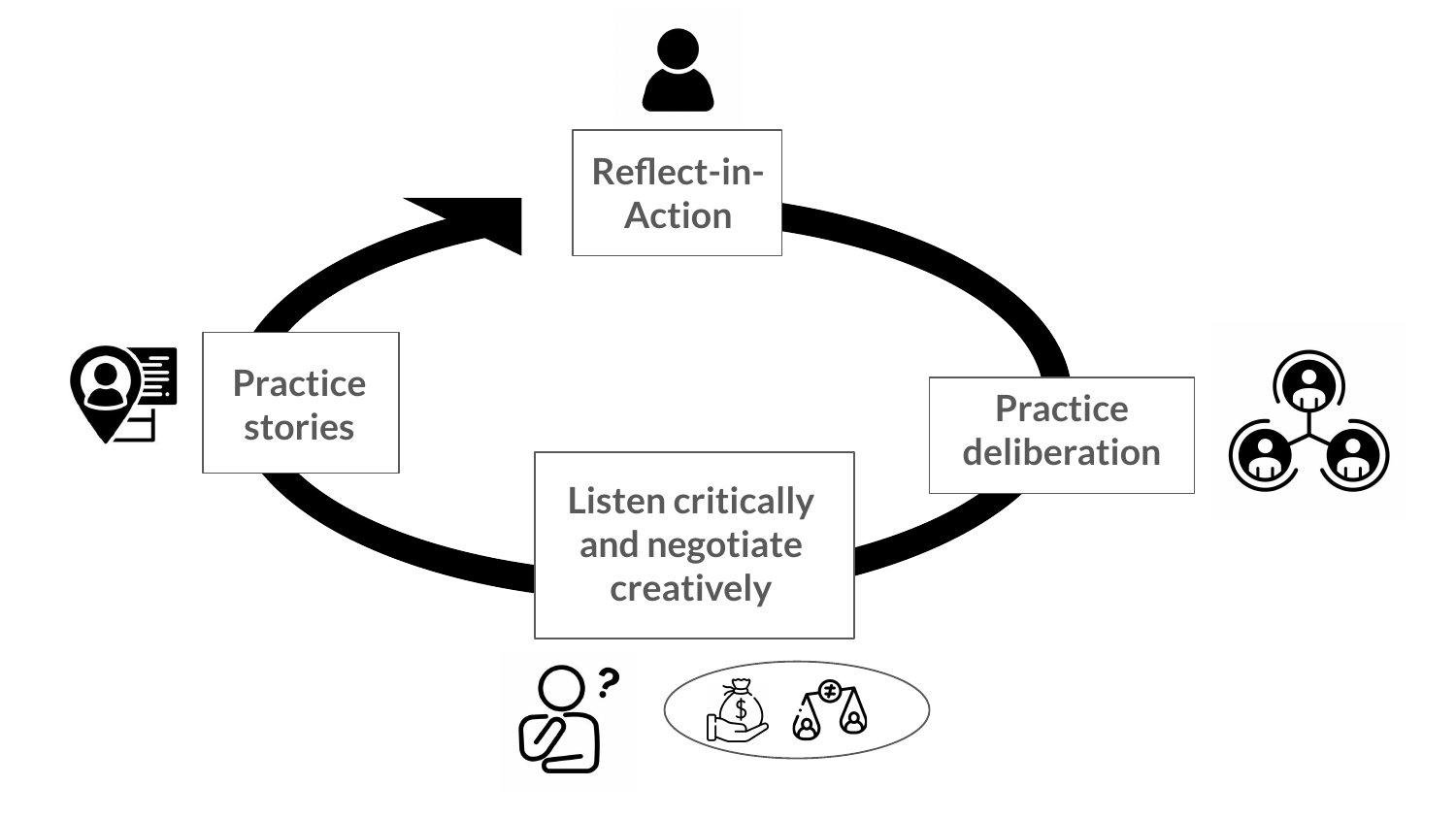}
  \caption{Presented here are the four elements of a critical pragmatism approach for algorithmic fairness: (1) reflect-in-action, (2) practice deliberation, (3) listen critically and negotiate creatively, and (4) share practice stories.}
  \label{fig:critical_pragmatism_framework}
\end{figure*}

In Figure~\ref{fig:critical_pragmatism_framework}, we visualize the four recommendations derived from critical pragmatism as a cycle, where recommendations are iterative and build on each other. This visualization underscores how our recommendations are not meant to act like a specific recipe or prescriptive set of steps, but rather offer a set of general-purpose guidelines to inform any aspect of the AI development lifecycle. Different recommendations may be more or less useful depending on context.

We then discuss how our recommendations might apply to specific stages in the AI development lifecycle, which provides more concrete guidance about how to advance critical pragmatism in practice. The specific stages are based on prior work related to pipeline-aware algorithmic fairness~\citep{black_toward_2023, suresh2021framework}.

\begin{itemize}
    \item \textbf{Viability:} \citet{black_toward_2023} define viability assessments as ``early investigations into whether including
an ML component within the decision-making system is preferable to the status quo.'' For viability assessments, \emph{practice stories} can be a useful starting point. Sharing practice stories might involve individual practitioners meeting prior to a project to discuss experiences from similar projects or an AI ethics class inviting speakers from the field to recount what they did in the past. This approach prioritizes first-hand accounts and personal experience in viability assessments, alongside other concerns about organizational capacity and practical constraints. Practice stories also ensure that mistakes learned in the past do not repeat. It may be clear immediately that a proposed AI tool will not work simply by listening to the experiences of one's colleagues.
    \item \textbf{Data collection:} In settings where data may not be sufficient, practitioners should think \emph{creatively} about how to collect or access the right data. This reflects our third recommendation: listen critically and negotiate creatively.   Embodying this recommendation might involve creative efforts to generate new data using qualitative data collection methods, surveys, or data scraping. The same is true in many planning processes, where data and information about a proposed planning action are sometimes only obtained through engaging with community groups and neighborhood residents directly (e.g., setting up a table at a local event or attending a public meeting).

    Separately, the data collection stage is a good time to set a precedent for communication (e.g., set up regular meetings) and consider which stakeholders should and need to be included. The urban planning literature has dealt extensively with decision-making around which stakeholders to include and how to structure effective stakeholder engagement processes. In addition to Forester's work discussed throughout this paper, we highlight a couple of examples. \textit{Structuring participatory processes:} \citet{Innes01012004} discuss the inadequacies of participation in the U.S. through processes like public hearings and notice-and-comment. They articulate a vision for the future involving collaborative participation and identify concrete next steps, such as shifting training resources and funding and changing government procedures. \textit{Deciding which stakeholders to include:}
    \citet{levine2021leadership}'s educational book teaching future planners about leadership is an example of  a useful framework. Categorizing stakeholders as  ``latents,'' ``apathetics,'' ``promoters,'' or ``defenders,'' Levine argues that stakeholders should be chosen strategically, based on their level of interest and influence.
    
    \item \textbf{Data pre-processing:} Reflection-in-action can improve data pre-processing, which is often iterative and incremental. For example, the downstream consequences of a decision about how to impute missing values or whether to transform input features may be uncertain. 
    Reflection-in-action encourages simulations, experimentation, and prototyping. Through reflection, practitioners may develop useful heuristics, based on past experience or the advice of colleagues.

\item \textbf{Deployment \& Integration:} To support the successful deployment of a tool and to integrate it into existing systems, engaging with stakeholders is critical. This involves actually listening to their concerns and approaching the problem collaboratively. We refer to the communication and stakeholder engagement processes set up in the data collection stage (discussed above). Lean on these existing processes. Ensure there are spaces for individuals to air their grievances and resolve concerns, prior to deploying a tool. Even after after initial deployment, continue to include stakeholders through working groups and advisory committees, which can encourage accountability and foster more collaborative participation.
\end{itemize}
\section{Addressing Failure Modes}
\label{sec:failure_modes}
In this section, we provide two brief responses to some concerns. The first is the concern that genuine participation, involving deliberation, may be slow and tends to preserve the status quo. This is a broader issue in urban planning that has faced widespread criticism.
The second is the concern that our recommendations are primarily incremental and may be of little use to individuals who lack power or the ability to shape decisions in environments where fairness and equity are low priority.
\paragraph{Concern 1: Burden on Time and Resources} Participatory processes can be slow. An example in urban planning is environmental review, which is typically required for major land use decisions~\citep{ceqr_explanation}. Environmental review can be lengthy, bureaucratic, and expensive. There are many points in the process where stakeholders may provide input. Sometimes this input is useful, but it can also slow down and even prevent meaningful improvements, such as the development of more affordable housing~\citep{scally2015democracy}. While critiques of participatory processes may be valid in some settings, we do not believe that they eliminate the need for participatory methods in all settings. Ignoring or eliminating opportunities for stakeholders to provide input may lead to additional obstacles later on, eradicate trust, and result in suboptimal outcomes. Instead, these critiques might indicate that we ultimately need to design better stakeholder engagement processes: for example, fixing the tendency to prioritize those with the most resources and time or finding ways to address stakeholders’ real concerns that may lie underneath seemingly surface-level conflicts. 
\paragraph{Concern 2: Lack of Decision-making Power} Our recommendations are still useful even in the kind of setting described here. Though it may be challenging to hold public meetings and practice deliberation, the advice to ``negotiate creatively’’ becomes even more pertinent. Instead of accepting the lack of power to change decisions, the critical pragmatist might still seek out a compromise solution. For example, even if the leadership in a company is indifferent to issues of fairness and equity, they may be motivated by other concerns (e.g., profit, customer satisfaction) that depend on building a fair and equitable product. Finding ways to develop ethical AI systems may still be possible through a careful and thoughtful re-framing  of the problem. 

Additionally, while we argue that public meetings support deliberation, there may be creative alternatives to still engage with stakeholders meaningfully and listen to their concerns: for example, conducting interviews with multiple participants, organizing a working group, or collecting stories of individuals' firsthand experiences. 

\onecolumn
\FloatBarrier
\section{Mapping ``Wicked Problems'' to Problems in Urban Planning and AI Ethics}
\label{sec:appendix_wicked_problems}

In this section, we describe how both an urban planning problem (affordable housing) and a problem in AI ethics (fair mortgage lending) map onto the ten criteria of wicked problems~\citep{rittel_dilemmas_1973}, formalizing the overlap between the two disciplines. At a high-level, we describe how affordable housing is a wicked problem in Section~\ref{fairness_planning}. We discuss the mortgage lending example in the first case study in Section~\ref{applying-pragmatism}.

\centering
\small{
\begin{longtable}{|p{6.5in}|}
\hline
\rule{0pt}{3ex}
      \textbf{(1) ``There is no definitive formulation."}
     \begin{itemize}[nosep]
         \item[{\includegraphics[width=1em]{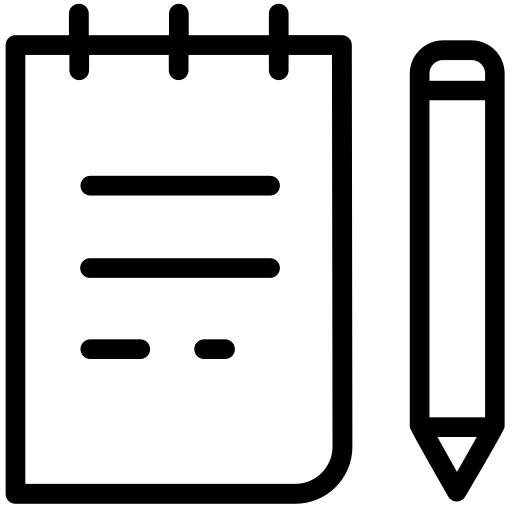}}] There are numerous explanations for why affordable housing is difficult to produce: governmental funding constraints, mismatched public and private interests, racial discrimination, redlining, zoning, and more. None of these perspectives solely explains the problem, and each suggests a slightly different approach to solve it.
         \item[{\includegraphics[width=1em]{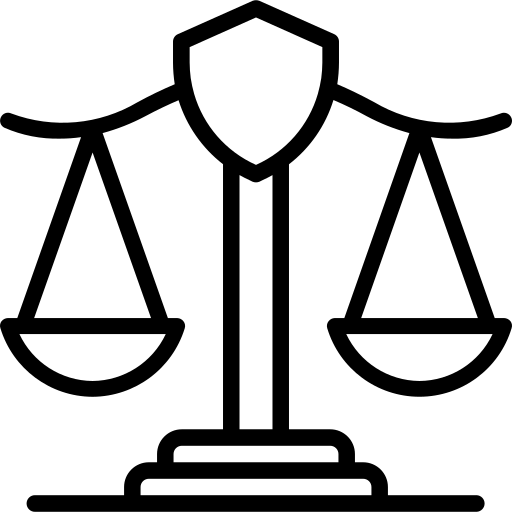}}] Differing perspectives on designing fair lending algorithms arise from the wickedness of the underlying problem (disparities in homeownership), disagreements about fairness definitions in housing contexts, and the validity of proposed implementations. For example, consider the fairness of a lending algorithm with respect to racial and ethnic gaps in homeownership. Some might consider the lending algorithm fair if predictions are the same for applicants with the same attributes aside from race. Others may argue for race corrections to rectify historical and social inequities in homeownership.
     \end{itemize} \\
     \hline
     \rule{0pt}{3ex}
   \textbf{(2) ``Wicked problems have no stopping rule."}
   \begin{itemize}[nosep]
   \item[{\includegraphics[width=1em]{diagrams/planning-icon.png}}] There is no point at which a policymaker can claim definitively to have solved the problem of providing affordable housing. Consider a program to produce more affordable housing using federal tax credits. Perhaps the program accomplishes its immediate goals. But, with new funding sources or more creative planning, the policymaker could potentially develop even more housing with better quality units or at lower cost.
       \item[{\includegraphics[width=1em]{diagrams/fairness.png}}] 
       Similarly, there are always ways that a lending algorithm can be \emph{more fair} depending on both the framing of the problem and the desired fairness outcomes. 
       Perhaps data pre-processing steps result in an algorithm with accurate, unbiased predictions. Some may consider the lending algorithm fair at this point. Still, racial and ethnic disparities in homeownership likely persist in society. Additional measures --- such as considering different risk thresholds for historically disadvantaged groups or implementing special purpose credit programs like down payment assistance \citep{mehrotra2024evidence, so2022beyond} --- may further mitigate these disparities.
   \end{itemize} \\
   \hline
   \rule{0pt}{3ex}
   \textbf{(3) ``Solutions ... are not true-or-false, but good-or-bad."}
   \begin{itemize}[nosep]
       \item[{\includegraphics[width=1em]{diagrams/planning-icon.png}}] There is no single \textit{correct} solution to providing affordable housing. Individuals--with diverse backgrounds, expertise, values, and ideologies--can evaluate potential solutions. Their judgments may not align, as they are based on perceptions of goodness rather than correctness.
       \item[{\includegraphics[width=1em]{diagrams/fairness.png}}] Anyone can assess whether an algorithm is fair, basing such judgments depend on individual values and experiences. For example, engineers may prefer technical debiasing measures while policymakers may seek improvements in concrete outcomes.
    \end{itemize} \\
    \hline
    \rule{0pt}{3ex}
    \textbf{(4) ``There is no immediate and no ultimate test of a solution to a wicked problem."}
   \begin{itemize}[nosep]
   \item[{\includegraphics[width=1em]{diagrams/planning-icon.png}}] It is difficult to evaluate the success of any solution since its impact can vary over time and some outcomes may be challenging to measure. For example, immediate issues might arise from a new policy: increased public expenditure or neighborhood opposition. At the same time, longer-term benefits may cancel out shorter-term issues--benefits that are harder to measure like improved housing stability, employment outcomes, and health.
       \item[{\includegraphics[width=1em]{diagrams/fairness.png}}] Isolating the impact of the lending algorithm may be difficult, particularly alongside other social processes and as effects accumulate over time. Additionally, the data used to test an algorithm at later time points may look very different from the initial training data as outcomes and behaviors change~\citep{koh2021wilds}. Such distribution shifts may be inevitable and hard to anticipate, as the case of Google Flu Trends demonstrated \citep{lazer2014parable}.
    \end{itemize} \\
    \hline
    \rule{0pt}{3ex}
   \textbf{(5) ``Every solution to a wicked problem is a `one-shot operation'; because there is no opportunity to learn by trial-and-error, every attempt counts significantly."}
   \begin{itemize}[nosep]
       \item[{\includegraphics[width=1em]{diagrams/planning-icon.png}}] Poorly designed housing policies can be challenging to undo and even irreversible. For example, unstable and high-cost housing is a risk factor for serious health complications, including death \citep{nyamathi2021impact, zivanovic2015impact, aldridge2018morbidity, meltzer2016housing}. Housing can impact wealth~\citep{rothstein2017color, goodman2018homeownership} and education~\citep{brennan2011impacts}. Considerable amounts of money and resources may be required (e.g., building new housing). As a result, policymakers cannot freely experiment with different policy approaches and choose the best one. There are severe consequences associated with implementing any of them. 
       \item[{\includegraphics[width=1em]{diagrams/fairness.png}}] Biased lending decisions can have similarly harmful and irreversible effects. There are the administrative costs associated with building and deploying the tool, and the social costs, like reduced housing stability and wealth building~\citep{markup_mortgage}. In turn, these can reinforce and even exacerbate the effects of poorly executed and discriminatory housing policy. 
    \end{itemize} \\
    \hline
    \rule{0pt}{3ex}
   \textbf{(6) ``Wicked problems do not have an enumerable (or an exhaustively describable) set of potential solutions, nor is there a well-described set of permissible operations that may be incorporated into the plan."}
   \begin{itemize}[nosep]
       \item[{\includegraphics[width=1em]{diagrams/planning-icon.png}}] Innovative, unexpected ideas may prove just as effective as traditional ones. Policymakers may turn to creative design solutions for building affordable housing. For example, building ``half a house" is one way to scale affordable housing at low cost while fostering personalization and independence \citep{aravena}. Alternative ownership models like housing cooperatives and land trusts also aim to provide affordable housing \citep{meehan2014reinventing}.
       \item[{\includegraphics[width=1em]{diagrams/fairness.png}}] There are both broad approaches to fair algorithm design, with foundations across many disciplines \citep{pessach2022review}, as well as research on algorithmic methods in mortgage lending specifically \citep{azizi2018designing, lee2021algorithmic, so2022beyond}. Ongoing research in these fields suggests the potential for new, creative solutions to fair lending algorithms, like incorporating the concept of reparations into loan denial decisions.
    \end{itemize} \\
    \hline
    \rule{0pt}{3ex}
   \textbf{(7) ``Every wicked problem is essentially unique."}
   \begin{itemize}[nosep]
       \item[{\includegraphics[width=1em]{diagrams/planning-icon.png}}] Housing policies that work in one location may not work elsewhere. The challenges of developing affordable housing vary based on local factors like politics, historical context, job markets, or land value.
       \item[{\includegraphics[width=1em]{diagrams/fairness.png}}] It may be possible to reuse elements of a prior approach, but factors like data availability, organizational dynamics, and stakeholder values mean that the problem will be unique. For example, race-blind algorithms can remove bias in some contexts, but in others, particularly when historical data already encode the same patterns, more extensive mitigation approaches may be necessary~\citep{barocas-hardt-narayanan}.
    \end{itemize} \\
    \hline
   \textbf{(8) ``Every wicked problem can be considered to be a symptom of another problem."}
   \begin{itemize}[nosep]
       \item[{\includegraphics[width=1em]{diagrams/planning-icon.png}}] Challenges with affordable housing are related to broader systemic and historical issues, such as income inequality, limited public funding for housing, and redlining.
       \item[{\includegraphics[width=1em]{diagrams/fairness.png}}] Reducing disparities in homeownership is similarly rooted in broader issues: the legacy of housing discrimination, predatory lending practices, and systemic barriers to wealth accumulation. These problems, in turn, contribute to bias at multiple stages in the design of mortgage lending algorithms --- e.g., in the training data or in the deployment of the tool.
    \end{itemize} \\
    \hline
    \rule{0pt}{3ex}
   \textbf{(9) ``The existence of a discrepancy representing a wicked problem can be explained in numerous ways. The choice of explanation determines the nature of the problem's resolution."}
   \begin{itemize}[nosep]
       \item[{\includegraphics[width=1em]{diagrams/planning-icon.png}}] In the sciences, there is often a single explanation of an unexpected result. But there are many ways to refute a wicked problem. For example, consider a program to develop subsidized housing units. If unsuccessful, policymakers might find there are multiple explanations: perhaps it did not lower rents for the populations most in need or it displaced existing affordable housing. 
       \item[{\includegraphics[width=1em]{diagrams/fairness.png}}] There may be multiple explanations for why an algorithm is unfair. People tend to pick the one that fits their ``world view" \citep{rittel_dilemmas_1973}. In the case of biased lending algorithms, the source of the bias could be explained by existing disparities in the data, the choice of variables used through feature selection methods, methods for data pre-processing and cleaning, technical decisions made during model development, and other changes that emerge through deployment~\citep{friedman1996bias}.
    \end{itemize} \\    \hline
    \rule{0pt}{3ex}
   \textbf{(10) ``The planner has no right to be wrong...Planners are liable for the consequences of the actions they generate."}
   \begin{itemize}[nosep]
       \item[{\includegraphics[width=1em]{diagrams/planning-icon.png}}] Developing affordable housing in the U.S. is critical. With limited resources and widespread need, policymakers cannot afford to be wrong. As a result, they may settle for incremental changes that improve on the status quo--e.g. expanding access to rental assistance or increasing public money for new housing. They would be unconvinced by universal solutions and wary of potential missteps that can cause harm to real people. 
       \item[{\includegraphics[width=1em]{diagrams/fairness.png}}] The stakes of deploying biased and harmful lending algorithms are high. Practitioners must keep in mind that their work will directly impact individuals and communities.
    \end{itemize} \\
    \hline
\caption{\textbf{Wicked Problems in Urban Planning \includegraphics[width=1em]{diagrams/planning-icon.png} and AI Ethics \includegraphics[width=1em]{diagrams/fairness.png}}. The bolded phrases are the 10 characteristics of wicked problems per \citet{rittel_dilemmas_1973}. The explanations below discuss the urban planning example from Section~\ref{fairness_planning} \includegraphics[width=1em]{diagrams/planning-icon.png} and an AI ethics example from the first case study in Section~\ref{applying-pragmatism}  \includegraphics[width=1em]{diagrams/fairness.png}.}
\label{tab:wicked-problems-ex}
\end{longtable}
}

\end{document}